\documentclass[fleqn,usenatbib]{mnras}
\usepackage{newtxtext,newtxmath}
\usepackage[T1]{fontenc}

\DeclareRobustCommand{\VAN}[3]{#2}
\let\VANthebibliography\thebibliography
\def\thebibliography{\DeclareRobustCommand{\VAN}[3]{##3}\VANthebibliography}

\usepackage{graphicx}
\usepackage{amsmath}

\usepackage{bm}
\usepackage{cancel}
\usepackage{epsfig}
\usepackage{dcolumn}
\usepackage{hhline}
\usepackage{enumerate}
\usepackage[utf8]{inputenc}
\usepackage[dvipsnames]{xcolor}
\usepackage{bm}
\usepackage{caption}
\usepackage{subcaption}
\usepackage[normalem]{ulem}

\usepackage{scalerel}
\usepackage{tikz}
\usetikzlibrary{svg.path}

\definecolor{orcidlogocol}{HTML}{A6CE39}
\tikzset{
  orcidlogo/.pic={
    \fill[orcidlogocol] svg{M256,128c0,70.7-57.3,128-128,128C57.3,256,0,198.7,0,128C0,57.3,57.3,0,128,0C198.7,0,256,57.3,256,128z};
    \fill[white] svg{M86.3,186.2H70.9V79.1h15.4v48.4V186.2z}
                 svg{M108.9,79.1h41.6c39.6,0,57,28.3,57,53.6c0,27.5-21.5,53.6-56.8,53.6h-41.8V79.1z M124.3,172.4h24.5c34.9,0,42.9-26.5,42.9-39.7c0-21.5-13.7-39.7-43.7-39.7h-23.7V172.4z}
                 svg{M88.7,56.8c0,5.5-4.5,10.1-10.1,10.1c-5.6,0-10.1-4.6-10.1-10.1c0-5.6,4.5-10.1,10.1-10.1C84.2,46.7,88.7,51.3,88.7,56.8z};
  }
}

\newcommand\orcidicon[1]{\href{https://orcid.org/#1}{\mbox{\scalerel*{
\begin{tikzpicture}[yscale=-1,transform shape]
\pic{orcidlogo};
\end{tikzpicture}
}{|}}}}

\def\apj{{Astroph.\@ J.\ }}

\def \beq  {\begin{equation}}
\def \eeq  {\end{equation}}
\def \ber  {\begin{eqnarray}}
\def \eer  {\end{eqnarray}}

\def \omm  {\Omega_{0 {\rm m}}}

\newcommand{\newc}{\newcommand}

\newc{\be}{\begin{equation}}
\newc{\ee}{\end{equation}}
\newc{\ba}{\begin{eqnarray}}
\newc{\ea}{\end{eqnarray}}
\newc{\bea}{\begin{eqnarray*}}
\newc{\eea}{\end{eqnarray*}}
\newc{\D}{\partial}
\newc{\ie}{{\it i.e.} }
\newc{\eg}{{\it e.g.} }
\newc{\etc}{{\it etc.} }
\newc{\etal}{{\it et al.}}
\newc{\lcdm}{$\Lambda$CDM }
\newc{\lcdmnospace}{$\Lambda$CDM}
\newc{\omom}{$\Omega_{0 \rm m}$ }
\newc{\omomnospace}{$\Omega_{0 \rm m}$}
\newc{\plcdm}{Planck/$\Lambda$CDM }
\newc{\plcdmnospace}{Planck/$\Lambda$CDM}
\newc{\calm}{$\cal{M}$ }
\newc{\calmnospace}{$\cal{M}$}

\newcommand{\nn}{\nonumber}
\newc{\ra}{\Rightarrow}

\title[Hints for possible low redshift oscillation around BF $\Lambda$CDM]{Hints for possible low redshift oscillation around the best fit $\Lambda$CDM model in the expansion history of the Universe}

\author[L. Kazantzidis et al.]{
L. Kazantzidis\orcidicon{0000-0002-7134-4324},$^{1}$\thanks{l.kazantzidis@uoi.gr}
H. Koo\orcidicon{0000-0003-0268-4488},$^{2,3}$\thanks{hkoo@kasi.re.kr}
S. Nesseris\orcidicon{0000-0002-0567-0324},$^{4}$\thanks{savvas.nesseris@csic.es}
L. Perivolaropoulos\orcidicon{0000-0001-9330-2371}$^{1}$\thanks{leandros@uoi.gr}
and A. Shafieloo\orcidicon{0000-0001-6815-0337},$^{2,3}$\thanks{shafieloo@kasi.re.kr}
\\
$^{1}$Department of Physics, University of Ioannina, GR-45110, Ioannina, Greece\\
$^{2}$Korea Astronomy and Space Science Institute, Daejeon 34055, Korea\\
$^{3}$University of Science and Technology, Yuseong-gu 217 Gajeong-ro, Daejeon 34113, Korea\\
$^{4}$Instituto de F\'isica Te\'orica UAM-CSIC, Universidad Auton\'oma de Madrid, Cantoblanco, 28049 Madrid, Spain
}

\date{Accepted 2020 December 8. Received 2020 December 7; in original form 2020 October 28}
\pubyear{2021}

\begin{document}

\label{firstpage}
\pagerange{\pageref{firstpage}--\pageref{lastpage}}

%\preprint{IFT-UAM/CSIC-20-138}
\maketitle

\begin{abstract}
We search for possible deviations from the expectations of the concordance $\Lambda$CDM model in the expansion history of the Universe by analysing the Pantheon Type Ia Supernovae (SnIa) compilation along with its Monte Carlo simulations using redshift binning. We demonstrate that the redshift binned best fit $\Lambda$CDM matter density parameter $\Omega_{0 {\rm m}}$ and the best fit effective absolute magnitude $\cal M$ oscillate about their full dataset best fit values with considerably large amplitudes. Using the full covariance matrix of the data taking into account systematic and statistical errors, we show that at the redshifts below $z\approx0.5$ such oscillations can only occur in 4 to 5$\%$ of the Monte Carlo simulations. While statistical fluctuations can be responsible for this apparent oscillation, we might have observed a hint for some behaviour beyond the expectations of the concordance model or a possible additional systematic in the data. If this apparent oscillation is not due to statistical or systematic effects, it could be due to either the presence of coherent inhomogeneities at low $z$ or due to oscillations of a quintessence scalar field.
\end{abstract}


\section{Introduction}

In 1998, two independent groups \cite{Riess:1998cb,Perlmutter:1998np} confirmed that the Universe is undergoing a phase of accelerated expansion, which has been attributed to the cosmological constant \cite{Carroll:2000fy}, thus establishing \lcdm as the concordance model of modern cosmology. Despite its simplicity and consistency with most cosmological observations for almost two decades \cite{Betoule:2014frx,Aubourg:2014yra,Baxter:2016ziy,Alam:2016hwk,Efstathiou:2017rgv,Scolnic:2017caz,Aghanim:2018eyx}, \lcdm faces some challenges at the theoretical level \cite{Weinberg:1988cp,Sahni:2002kh,steinhardtbook,Velten:2014nra}, as well as at the observational one, since recent observations revealed some inconsistencies between the measured values of the basic parameters of \lcdm \cite{Sahni:2014ooa,Sola:2016jky,Zhao:2017cud,DiValentino:2019qzk,Handley:2019tkm,Li:2019san,Arjona:2019fwb,Arjona:2020kco}. 

The most prominent tension in the context of \lcdm is the so-called ``$H_0$ tension", which describes the discrepancy between the Planck indirect measurement of the Hubble parameter $H_0$, from Cosmic Microwave Background (CMB), Baryon Acoustic Oscillations (BAO) and uncalibrated Type Ia supernovae (SnIa) data using the inverse distance ladder method \cite{Aghanim:2018eyx} with the direct measurement published from  SnIa data, using the standard distance ladder method (\ie calibrated SnIa \cite{Riess:2019cxk,Riess:2020sih}). This discrepancy is currently at a $4.4 \sigma$ level. Moreover, a tension that is currently at a $2-3 \sigma$ level,  is the so-called ``growth tension", which refers to the mismatch between the $\sigma_8$ (density rms matter fluctuations in spheres of radius of about $8 \, h^{-1} \textrm{Mpc}$) and/or \omom (matter density parameter) measurement of the Planck mission \cite{Aghanim:2018eyx} with Weak Lensing (WL) \cite{Hildebrandt:2016iqg,Kohlinger:2017sxk,Joudaki:2017zdt,Abbott:2017wau,Heymans:2020gsg}, Redshift Space Distortion (RSD) data \cite{Macaulay:2013swa,Sola:2016zeg,Basilakos:2017rgc,Nesseris:2017vor,Kazantzidis:2018rnb,Perivolaropoulos:2019vkb,Kazantzidis:2019dvk,Skara:2019usd}. as well as cluster count data (which report consistently lower values of $\sigma_8$) \cite{Bohringer:2014ooa,Ade:2015fva,deHaan:2016qvy}

In order to explain the aforementioned challenges a plethora of theories have been proposed in the literature to solve the theoretical \cite{ArmendarizPicon:2000dh,Zimdahl:2000zm,Moffat:2005ii,Grande:2006nn,CalderaCabral:2008bx,Benisty:2018qed,Anagnostopoulos:2019myt} and the observational challenges of \lcdmnospace. In particular, for the observational challenges the mechanisms that have been proposed and can alleviate one or even both of these tensions simultaneously include early \cite{Karwal:2016vyq,Hazra:2018opk,Poulin:2018cxd,Agrawal:2019lmo,Keeley:2020rmo} and late dark energy models \cite{DiValentino:2017zyq,Yang:2018qmz,Yang:2019jwn,Li:2019yem,Vagnozzi:2019ezj,Li:2020ybr,Alestas:2020mvb}, interacting dark energy models \cite{Yang:2018euj,Yang:2018uae,DiValentino:2019ffd,DiValentino:2019jae,Lucca:2020zjb,Gomez-Valent:2020mqn}, metastable dark energy models \cite{Shafieloo:2016bpk,Szydlowski:2018kbk,Li:2019san,Yang:2020zuk}, modified gravity theories \cite{Ballardini:2016cvy,Lin:2018nxe,Rossi:2019lgt,Escamilla-Rivera:2019ulu,Braglia:2020iik,Kazantzidis:2020tko,Ballesteros:2020sik} as well as modifications of the basic assumptions of \lcdm such as non zero spatial curvature \cite{Ooba:2017ukj,Park:2017xbl}, and many more \cite{Joudaki:2016kym,Zhao:2017urm,Zhao:2017urm,Sola:2017znb, Gomez-Valent:2018nib,Colgain:2019pck,MPutten:2019,Camarena:2019moy} (see also the reviews \cite{Huterer:2017buf,Ishak:2018his,Kazantzidis:2019dvk} and references within). 

\begin{table*}
\caption{The best fit values with the $1 \sigma$ error of \calm and \omom for the four redshift bins with equal number of datapoints for the real data. Notice that for first three redshift bins the $\sigma$ distance ($\Delta \sigma$) of the best fit from the full dataset best fit is at least $1\sigma$ and on the average it is larger than $1.2\sigma$. In the simulated Pantheon data such large simultaneous deviations for the first three bins occurs for about $2\%$ of the datasets. }
\label{tab:redtom}
\begin{centering}
\begin{tabular}{|c|c|c|c|c|c|c}
 \hline 
 \rule{0pt}{3ex}  
  Bin & $z$ Range & ${\cal{M}} \pm 1 \sigma \text{ error}$ & $\Delta \sigma_{{\cal{M}}}$ & $\omm \pm 1 \sigma \text{ error}$ & $\Delta \sigma_{\Omega_{0 \rm m}}$ \\
    \hline
    \rule{0pt}{3ex}  
Full Data  & $0.01<z<2.26$ & $23.81 \pm 0.01$  & - & $0.29 \pm 0.02$ & -\\    
1st & $0.01<z<0.13$ & $23.78 \pm 0.03$  & 1.14 & $0.07 \pm 0.17$ & 1.35 \\
2nd & $0.13<z<0.25$ & $23.89 \pm 0.06$  & 1.48  & $0.56 \pm 0.19$ & 1.34\\
3rd & $0.25<z<0.42$ & $23.75 \pm 0.06$  & 0.99 & $0.18 \pm 0.11$ & 1.05\\
4th & $0.42<z<2.26$ & $23.85 \pm 0.06$  & 0.69  & $0.33 \pm 0.06$ & 0.50 \\
\hline
\end{tabular}
\end{centering}
\end{table*}

The measurement of $H_0$ that has been published by the SnIa data leading to the ``$H_0$ tension" is based on the assumption that SnIa can be considered as standard candles, thus allowing to probe the Hubble parameter through the apparent magnitude
\be 
m(z)=M+5 \log_{10} \left[\frac{d_L (z)}{1 \textrm{Mpc}} \right]+25, \label{eq:mbth}
\ee
where $d_L(z)$ is the luminosity distance, which in a flat Universe can be expressed as 
\be 
d_L(z)=c (1+z) \int_0^z \frac{dz'}{H(z')}, \label{eq:dlflat}
\ee
while $M$ corresponds to the corrected, over stretch and color, absolute magnitude. 

Alternatively, the apparent magnitude can be expressed in terms of the dimensionless Hubble-free luminosity distance $D_L \equiv H_0 \, d_L/c$ as
\be 
m(z)=M +5 \log_{10}\left[D_L(z)\right] + 5 \log_{10}\left(\frac{c/H_0}{1\textrm{Mpc}}\right)+25. \label{eq:mbthdim}
\ee
Clearly, from Eq. \eqref{eq:mbthdim} it is evident that the parameters $H_0$ and $M$ are degenerate and since in the context of \lcdm both of these are assumed to be constant, usually, a marginalization process is performed \cite{Conley:2011ku,Betoule:2014frx,Scolnic:2017caz} over the degenerate combination 
\ba 
{\cal M} &\equiv& M +5 \log_{10}\left[\frac{c/H_0}{1 \textrm{Mpc}}\right]+25\nn \\
&=& M -5 \log_{10}(h)+42.38, \label{eq:calmdef}
\ea
where $h \equiv H_0/100 \;\textrm{km} \, \textrm{s}^{-1}\, \textrm{Mpc}^{-1}$. However, in our analysis we choose to keep \calm in order to avoid any loss of crucial information. 

The latest (and largest thus far) compilation of SnIa that has been published is the Pantheon dataset \cite{Scolnic:2017caz}, consisting of 1048 SnIa in the redshift range $0.01<z<2.3$. Using Eqs.~\eqref{eq:mbth}-\eqref{eq:calmdef}, the corresponding $\chi^2$ function reads
\be 
\chi^2 ({\cal M}, \Omega_{0 \rm m})=V^i_\textrm{SnIa} \, C_{ij}^{-1} \, V^j_\textrm{SnIa}, \label{chifunc}
\ee  
where $V^i_\textrm{SnIa}\equiv m_{obs}(z_i)-m(z)$ and $C_{ij}^{-1}$ is the inverse covariance matrix. The covariance matrix can be considered as the sum of two matrices: a diagonal matrix that is associated with the statistical uncertainties of the  apparent magnitude $m_{obs}$ of each SnIa and a non-diagonal part that is connected with the systematic uncertainties due to the bias correction method \cite{Scolnic:2017caz}.

In Refs.~\cite{Kazantzidis:2020tko,Sapone:2020wwz} it was shown that the best fit \lcdm parameter values for the best fit parameters \calm and \omom of redshift binned Pantheon data oscillate around the full dataset best fit at a level that is consistently larger than $1\sigma$ for the first three out of four redshift bins. A similar effect was observed in \cite{Colgain:2019pck}, where the best fit values of $\Omega_{0 {\rm m}}$ and $h$ for various maximum redshift cutoffs $z_{max}$ were studied instead. Here we should emphasize that any realization of a data based on a given model would have its own specific features and characteristics that might look unusual but in reality they are effects of random fluctuations. So to assign statistical significance to unusual features or behaviors seen in a given data (to evaluate if they can be real) it is necessary to compare the real data with a large number of Monte Carlo simulations. This comparison with 1000 random Monte Carlo simulations is a key part of our current analysis to evaluate how statistically significant such variations are.

If the variation we see in the real data is due to statistical fluctuations, then the same variation is anticipated to be evident in simulated Pantheon-like datasets. In this analysis we will address the following questions:
\begin{itemize}
\item How likely is this behaviour of the data in the context of the \lcdm model?
\item In how many realizations we can see more than the $\sigma$ deviations of the real data $(\sigma^{real})$ for both \calm and \omom in the first three or any three out of four redshift bins?
\item In how many realizations we can see more than the $1 \sigma$ deviations for both \calm and \omom in the first three or in any three out of four redshift bins?
\end{itemize}
The structure of the paper is the following: In Section~\ref{sec:mcmc} we describe the statistical analysis and the comparison of the constructed simulated datasets with the actual Pantheon data searching for abnormalities of the real data in the context of the reported level of Gaussian uncertainties. Finally, in Section~\ref{sec:conclusions} we summarize our results and discuss possible extensions of the present analysis.

\section{Real versus Monte Carlo Data}\label{sec:mcmc}
In our Monte Carlo statistical analysis we split the Pantheon dataset \cite{Scolnic:2017caz} into four redshift bins, consisting of equal number of datapoints (262). The number of bins is an important implicit parameter that could affect the results of our analysis. Too many bins may lead to overfiting of the data, while a very small number of bins may miss interesting signals hidden. In the present analysis we have chosen to use four bins with equal number of data. However, this choice is clearly not unique. For example, bins could have been chosen so that each bin has the same redshift interval, while it is not appropriate for the present analysis since most of SnIa in Pantheon are concentrated in the lower part of their redshift range, or has the same cumulative signal to noise (S/N) (including downweighting from systematics, which correlate points within the same bin). In fact an interesting extension of the present analysis would be the effect of the binning method on the strength and the statistical significance of the identified oscillating signal.

\begin{figure*}
\centering
\includegraphics[width = 1.0 \textwidth]{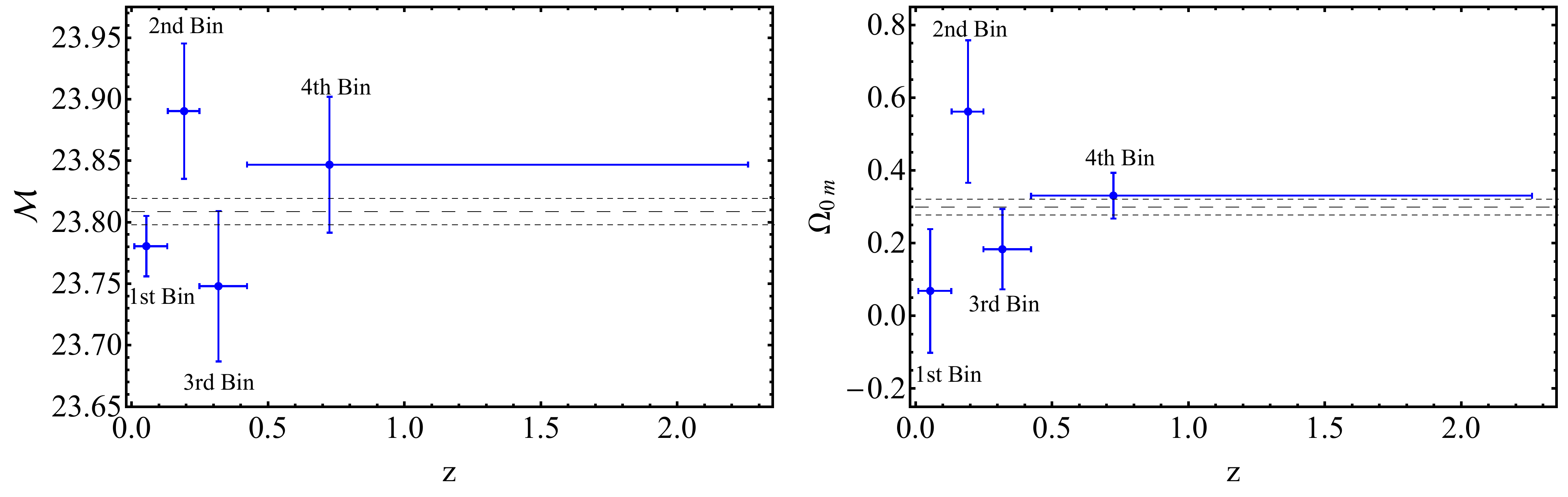}
\caption{The $1\sigma$ best fit values (blue dots) of $\cal{M}$ and $\Omega_{0 {\rm m}}$ of the real data for each bin. The dashed line corresponds to the best fit value of the full dataset, while the dot dashed line to its $1\sigma$ error. Clearly, the first three bins best fit differ by at least $1\sigma$ from the best fit values of the full dataset.}
\label{fig:crossplot}
\end{figure*}

We then find the best fit parameters \calm and \omom and $1\sigma$ uncertainties in the context of a \lcdm model for each bin, with \calm and \omom  being allowed to vary simultaneously. We also find the corresponding best fit for the full Pantheon dataset and identify the $\sigma$ distance between the best fit parameter values in each bin and the best fit value of the full dataset. The results of the tomography for the real data can be seen in Table \ref{tab:redtom} as well as in Fig. \ref{fig:crossplot}. Clearly, all first three bins of the real data best fits of \calm and \omom differ  by at least $1 \sigma$ from the full dataset best fits. 

\begin{figure*}
\centering
\includegraphics[width = 1.0 \textwidth]{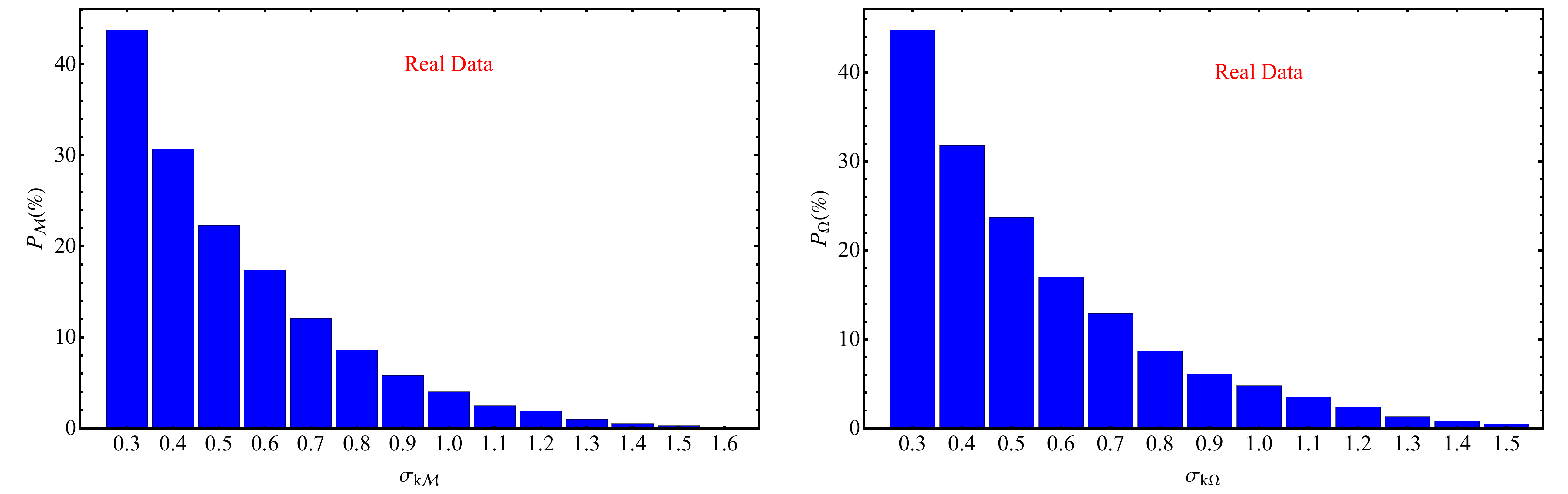}
\caption{\textit{Left-hand panel:} Percent of simulated Pantheon dataset (including systematics) where the first three out of four bins all differ simultaneously more than $\sigma_{k {\cal{M}}}\; \sigma$ from the best fit of the full dataset. The red dotted line corresponds to the real data that differ more than $1 \sigma$ ($1.14, 1.48$ and $0.99 \sigma$ for the first three bins respectively) from the full dataset best fits. \textit{Right-hand panel:} Same as the left-hand panel for the parameter \omom instead of \calmnospace.}
\label{fig:prob1}
\end{figure*} 

In order to estimate the likelihood of such a $\sigma$ deviation of best fit values in the first three bins, we construct 1000 simulated Pantheon-like datasets, with random apparent magnitudes $m$ obtained from a multivariate normal distribution with a mean value equal to the best fit \lcdm value of the real data using the full covariance matrix of the real data. 
The corresponding probability distribution is of the form
\be
f_{\bm{m}} \left(m_1, \ldots, m_k \right)=\frac{\exp \left[-\frac{1}{2} \left(\bm{m}-\bar{\bm{m}} \right)^T \bm{C}^{-1} \left(\bm{m}-\bar{\bm{m}} \right) \right]}{ \sqrt{(2 \pi)^k \, |\bm{C}|}},
\label{mvardist}
\ee
where $\bm{C}$ is the full non-diagonal covariance matrix including both statistical and systematic errors, $\bm{m}$ is the vector $\lbrace m_1,m_2, \ldots m_k \rbrace$ and $\bar{\bm{m}}$ corresponds to the mean value of the apparent magnitude vector.
Using this multivariate normal distribution we construct the simulated datasets and find the percent fraction of them where all first three redshift bins have best fit \lcdm parameter values  \calm and \omom that have {\it simultaneously} $\sigma$ distance from the real data best fit more than $k\; \sigma\equiv \sigma_k \; \sigma$. These results for the parameters  $\cal{M}$ ($\sigma_k=\sigma_{k {\cal{M}}})$ and  \omom ($\sigma_k=\sigma_{k \Omega})$ are shown in Fig. \ref{fig:prob1}.

\begin{figure*}
\centering
\includegraphics[width = 1.0 \textwidth]{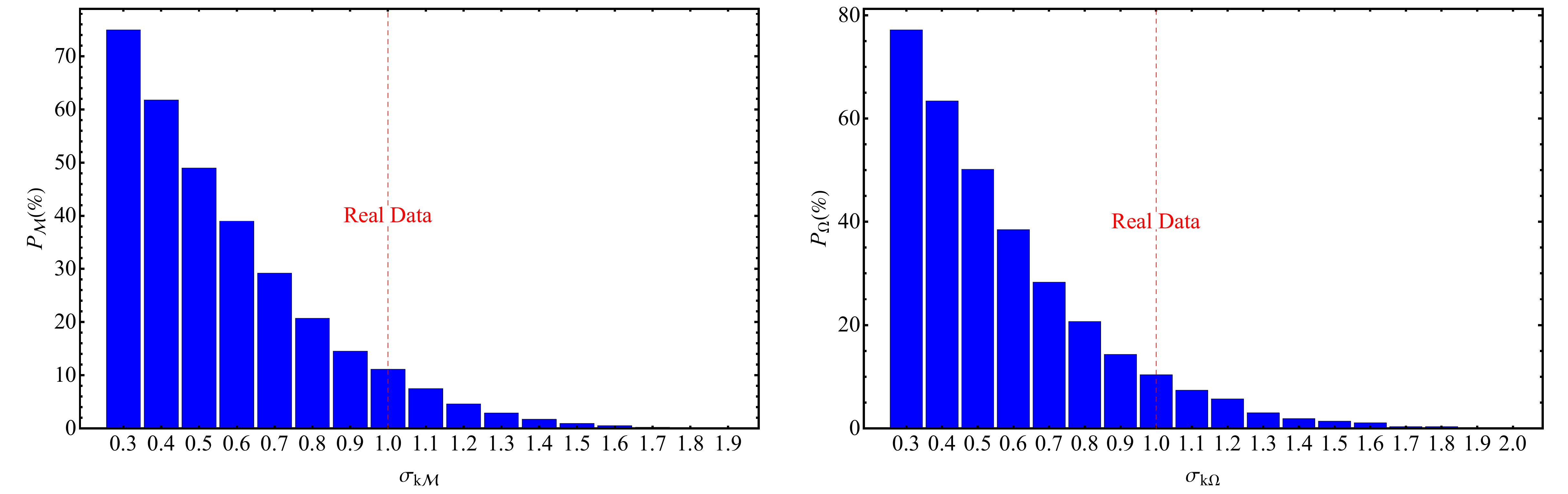}
\caption{\textit{Left-hand panel:} Percent of simulated Pantheon dataset (including systematics) where any three out of four bins all differ simultaneously more than $\sigma_{k {\cal{M}}}\; \sigma$ from the best fit of the full dataset. \textit{Right-hand panel:} Same as the left-hand panel for the parameter \omom instead of \calmnospace.}
\label{fig:prob2}
\end{figure*} 

\begin{table*}
\caption{Summary of the Monte Carlo deviations from the simulated and real data. The obtained results considering the exact $\sigma_{real}$ differences should be treated with care, since this decrease of the probabilities is not generic and it is based on a fine tuned $\sigma$ value.}
\label{tab:randres}
\begin{centering}
\begin{tabular}{|c|c|}
 \hline 
 \rule{0pt}{3ex}  
  Number of cases  & Probability \\
    \hline
    \rule{0pt}{3ex}  
\omom in the first 3 bins $> 1\sigma$ away from the best fit \omom to the whole data sample  & $ 4.8 \pm 2 \%$ \\
\calm in the first 3 bins $> 1 \sigma$ away from the best fit \calm to the whole data sample & $4 \pm 2.5 \% $ \\
 \omom in any 3 bins $> 1 \sigma$ away from the best fit \omom to the whole data sample & $ 10.4 \pm 2.2 \% $ \\
 \calm in any 3 bins $> 1 \sigma$  away from the best fit \calm to the whole data sample & $ 11.1 \pm 2.4 \% $\\
  \omom in the first 3 bins $>  \sigma^{real}$ away from the best fit \omom to the whole data sample & $1.4  \pm 2 \% $\\
  \calm in the first 3 bins $>  \sigma^{real}$ away from the best fit \calm to the whole data sample & $1.3 \pm 0.7 \% $\\
  \omom in any 3 bins $>  \sigma^{real}$ away from the best fit \omom to the whole data sample & $7.5 \pm 1.5 \% $\\
  \calm in any 3 bins $>  \sigma^{real}$ away from the best fit \calm to the whole data sample & $ 7.4 \pm 1.5 \% $ \\
\hline
\end{tabular}
\end{centering}
\end{table*}

According to Fig. \ref{fig:prob1}, the probability that all three first bins differ simultaneously more than $1 \sigma$ from the best fit of each simulated full dataset in the context of \lcdm is less than $5\%$. This is an effect approximately at $2\sigma$ level.

In fact, this probability is even smaller if we consider the exact $\sigma$ differences that are shown in Table \ref{tab:redtom} and find the fraction of simulated datasets with simultaneous $\sigma$ differences larger that the {\it exact} corresponding $\sigma$ differences of the real data. In particular we find that the probability to have simultaneously $1.14 \sigma$ difference (or larger) in the first bin, $1.48 \sigma $ difference (or larger) in the second bin and $0.99 \sigma$ difference (or larger) in the third bin for  \calmnospace, is $1.3 \pm 0.7 \%$. Similarly, for \omom we find the same probability to be $1.4  \pm 2 \%$. Even though this decrease of probability is interesting to note, it is not generic as it is based on the fine tuned $\sigma$ deviations of the real data bins from the full data best fits ($1.14 \sigma$, $1.48 \sigma$ and $0.99 \sigma$). Thus, this is an aposteriori statistic constructed after looking at the data.

Therefore, we adopt the more generic and conservative statistical level of significance of $5\%$ corresponding to the simultaneous deviation of at least $1\sigma$ for all three lowest $z$ bins. Note that a similar oscillating effect was also observed in Refs. \cite{Kazantzidis:2020tko,Sapone:2020wwz} even though its statistical significance was not quantified using simulated data as in the present analysis.

Moreover, it is interesting to check if this behaviour is also evident for {\it any} three out of four bins. In 1000 Monte Carlo realizations we find that the  number of simulated datasets where the derived  \omom in any 3 bins is more than $1 \sigma$ away from the best fit \omom to the whole (random) data sample is $10.4 \pm 2.2 \% $ while the corresponding number of cases for \calm is $ 11.1 \pm 2.4 \%$ as it is demonstrated in Fig. \ref{fig:prob2}. The probability is smaller if we consider the exact $\sigma$ difference of Table \ref{tab:redtom}. In particular, we derive the number of cases where the derived  \omom in any 3 bins is more than $\sigma^{real}_{\Omega_{0 \rm m}}$ away from the best fit \omom to the whole (random) data sample is $7.5 \pm 1.5 \% $, while the corresponding number of cases for \calm is $ 7.4 \pm 1.5 \%$. A summary of the results can be seen in Table \ref{tab:randres}. These results indicate that the aforementioned oscillating effect is much more prominent at low $z\lesssim 0.5$ where the dark energy density is more prominent than in the fourth bin, which involves higher $z$. This fact favors the possibility that the effect has a physical origin since a systematic effect would probably affect equally all four redshift bins.

\section{Conclusion - Outlook}\label{sec:conclusions}

We performed a redshift  tomography of the Pantheon data dividing them into four redshift bins of equal number of datapoints and searched for hints of abnormal oscillation behaviour for the best fit parameter values of \calm and \omom in these bins with respect to the corresponding best fits of the full Pantheon dataset. 

We constructed 1000 simulated Pantheon-like datasets and found that including both systematic and statistical uncertainties, the percentage of the simulated Pantheon dataset with a similar amplitude oscillating behaviour is $\simeq 5\%$. Considering only statistical uncertainties in the construction of the simulated datasets this probability decreases to about $2.7\%$.
 
While the statistical significance of the oscillations reduces when we consider any 3 bins out of 4 bins, we emphasise that the first three bins covering the $75\%$ of the total data points are all at relatively low redshifts ($z<0.42$) where dark energy is dominant. Hence, concerning the physical origin of the aforementioned effect, we anticipate that the importance of the first three bins is amplified compared to any other three bin combination.

The important issue here is how generic is the identified effect and also if it would have been expected in the context of a particular physical context. We argue that larger than expected oscillations around the standard model is a simple generic effect, especially if it is prominent at low redshifts where the effects of dark energy are more important. Thus, even though the look-elsewhere effect is hard to quantify in the context of the Monte-Carlo statistical analysis, the generic nature of the oscillating effect as well as the fact that it is more prominent at low $z$ where dark energy dominates, indicates that the statistical significance of the identified signal will not be significantly affected by the look elsewhere effect.

Plausible physical causes for such low $z$ oscillating behavior of the data include the following
\begin{itemize}
    \item The presence of large scale inhomogeneities at low $z$ including voids or superclusters \cite{Grande:2011hm,Shanks:2018rka}.
    \item Dark energy with oscillating density in redshift \cite{Xia:2006rr,Lazkoz:2010gz,Pace:2011kb,DeFelice:2012vd,Pan:2017zoh}. Such oscillations may be induced \eg by scalar field potentials with a local minimum \cite{Cicoli:2018kdo,Ruchika:2020avj}.
\end{itemize}

Finally, some interesting extensions of the present analysis include the following
 \begin{itemize}
     \item Further investigation for a similar oscillating behaviour in other data [\eg BAO or $H(z)$ cosmic chronometer data \cite{Marcondes:2017vjw,Ishak:2018his,Raveri:2019mxg}]. Regarding the cosmic chronometer data \cite{Marcondes:2017vjw}, even though no oscillating signal is evident, the errors are significantly larger than other probes and could well hide any interesting signal evident in other higher quality probes. On the contrary, regarding the BAO data \cite{Raveri:2019mxg} an interesting descending trend is evident, which could be interpreted as a hint for oscillations. Clearly, if such oscillations are observed in other cosmological datasets, the overall statistical significance of such an effect would be considerably boosted.
     \item Construction of physical models that naturally lead to such an oscillating low $z$ behavior of the data.
     \item Forecasts with future SnIa compilations, \eg by the LSST survey, to ascertain whether this oscillatory effect would be more prominent in upcoming data.
     \item Making some internal consistency checks such as using ``Robustness'' criterion \cite{Amendola:2012wc} or/and looking for redshift evolution in the light curve parameters of the data \cite{Koo:2020ssl} to determine whether the Pantheon sample is statistically consistent or is contaminated with systematics. 
 \end{itemize}

\section*{Acknowledgements}
The research of LK is co-financed by Greece and the European Union (European Social Fund- ESF) through the Operational Programme ``Human Resources Development, Education and Lifelong Learning" in the context of the project ``Strengthening Human Resources Research Potential via Doctorate Research – 2nd Cycle" (MIS-5000432), implemented by the State Scholarships Foundation (IKY). The research of LP is co-financed by Greece and the ESF through the Operational Programme  ``Human Resources Development, Education and Lifelong Learning 2014-2020" in the context of the project No. MIS 5047648. SN acknowledges support from the Research Projects PGC2018-094773-B-C32, the Centro de Excelencia Severo Ochoa Program SEV-2016-0597 and the Ram\'{o}n y Cajal program through Grant No. RYC-2014-15843. AS would like to acknowledge the support of the Korea Institute for Advanced Study (KIAS) grant funded by the government of Korea.

\section*{Data Availability}

The data access to the Pantheon compilation of SnIa is provided by \url{https://github.com/dscolnic/Pantheon}. Description of the Pantheon compilation is in \url{https://archive.stsci.edu/prepds/ps1cosmo/index.html} and \cite{Scolnic:2017caz}.

%\raggedleft
\bibliographystyle{mnras}
\bibliography{Bibliography}

\bsp
\label{lastpage}
\end{document}